\documentclass[twoside,twocolumn,9pt]{article}
\usepackage{extsizes}
\usepackage[super,sort&compress,comma]{natbib} 
\usepackage[version=3]{mhchem}
\usepackage[left=1.5cm, right=1.5cm, top=1.785cm, bottom=2.0cm]{geometry}
\usepackage{balance}
\usepackage{times,mathptmx}
\usepackage{sectsty}
\usepackage{graphicx} 
\usepackage{lastpage}
\usepackage[format=plain,justification=raggedright,singlelinecheck=false,font={stretch=1.125,small,sf},labelfont=bf,labelsep=space]{caption}
\usepackage{float}
\usepackage{fancyhdr}
\usepackage{fnpos}
\usepackage[english]{babel}
\usepackage{array}
\usepackage{droidsans}
\usepackage{charter}
\usepackage[T1]{fontenc}
\usepackage[usenames,dvipsnames]{xcolor}
\usepackage{setspace}
\usepackage[compact]{titlesec}

\usepackage{bm}

\usepackage{epstopdf}

\definecolor{cream}{RGB}{222,217,201}

\begin{document}

\pagestyle{fancy}
\thispagestyle{plain}


\makeFNbottom
\makeatletter
\renewcommand\LARGE{\@setfontsize\LARGE{15pt}{17}}
\renewcommand\Large{\@setfontsize\Large{12pt}{14}}
\renewcommand\large{\@setfontsize\large{10pt}{12}}
\renewcommand\footnotesize{\@setfontsize\footnotesize{7pt}{10}}
\makeatother

\renewcommand{\thefootnote}{\fnsymbol{footnote}}
\renewcommand\footnoterule{\vspace*{1pt}%
\color{cream}\hrule width 3.5in height 0.4pt \color{black}\vspace*{5pt}} 
\setcounter{secnumdepth}{5}

\makeatletter 
\renewcommand\@biblabel[1]{#1}            
\renewcommand\@makefntext[1]%
{\noindent\makebox[0pt][r]{\@thefnmark\,}#1}
\makeatother 
\renewcommand{\figurename}{\small{Fig.}~}
\sectionfont{\sffamily\Large}
\subsectionfont{\normalsize}
\subsubsectionfont{\bf}
\setstretch{1.125} 
\setlength{\skip\footins}{0.8cm}
\setlength{\footnotesep}{0.25cm}
\setlength{\jot}{10pt}
\titlespacing*{\section}{0pt}{4pt}{4pt}
\titlespacing*{\subsection}{0pt}{15pt}{1pt}

\renewcommand{\headrulewidth}{0pt} 
\renewcommand{\footrulewidth}{0pt}
\setlength{\arrayrulewidth}{1pt}
\setlength{\columnsep}{6.5mm}
\setlength\bibsep{1pt}

\makeatletter 
\newlength{\figrulesep} 
\setlength{\figrulesep}{0.5\textfloatsep} 

\newcommand{\topfigrule}{\vspace*{-1pt}%
\noindent{\color{cream}\rule[-\figrulesep]{\columnwidth}{1.5pt}} }

\newcommand{\botfigrule}{\vspace*{-2pt}%
\noindent{\color{cream}\rule[\figrulesep]{\columnwidth}{1.5pt}} }

\newcommand{\dblfigrule}{\vspace*{-1pt}%
\noindent{\color{cream}\rule[-\figrulesep]{\textwidth}{1.5pt}} }

\makeatother

\twocolumn[
  \begin{@twocolumnfalse}
\vspace{3cm}
\sffamily
\begin{tabular}{m{4.5cm} p{13.5cm} }

 &
\noindent\LARGE{\textbf{Atomic layer doping of Mn magnetic impurities from surface chains at a Ge/Si hetero-interface$^\dag$}} \\
\vspace{0.3cm} & \vspace{0.3cm} \\

 & \noindent\large{Koichi Murata,\textit{$^{a, b, \ast, \ddag, \textsection}$} Christopher Kirkham,\textit{$^{c, d,\ddag}$} Satoshi Tsubomatsu,\textit{$^{a, b}$} Takashi Kanazawa,\textit{$^{a, b}$} Kiyofumi Nitta,\textit{$^{e}$} Yasuko Terada,\textit{$^{e}$} Tomoya Uruga,\textit{$^{e}$} Koh-ichi Nittoh,\textit{$^{a}$} David R. Bowler,\textit{$^{c,f}$} and Kazushi Miki,\textit{$^{a, b, g, \ast}$}} \\

 &
\noindent\normalsize{We realize Mn $\delta$-doping into Si and Si/Ge interfaces using Mn atomic chains on Si(001). Highly sensitive X-ray absorption fine structure techniques reveal that encapsulation at room temperature prevents the formation of silicides / germanides whilst maintaining one dimensional anisotropic structures. This is revealed by studying both the incident X-ray polarization dependence and post-annealing effects. Density functional theory calculations suggest that Mn atoms are located at substitutional sites, and show good agreement with experiment. A comprehensive magnetotransport study reveals magnetic ordering within the Mn $\delta$-doped layer, which is present at around 120\,K. We demonstrate that doping methods based on the burial of surface nanostructures allows for the realization of systems for which conventional doping methods fail.} \\

\end{tabular}

\end{@twocolumnfalse} \vspace{0.6cm}

]

\renewcommand*\rmdefault{bch}\normalfont\upshape
\rmfamily
\section*{}
\vspace{-1cm}


\footnotetext{\textit{$^{a}$~National Institute for Materials Science (NIMS), Namiki 1-1, Tsukuba, 305-0044, Japan. E-mail: m-koichi@criepi.denken.or.jp, miki@eng.u-hyogo.ac.jp}}
\footnotetext{\textit{$^{b}$~Faculty of Pure and Applied Sciences, University of Tsukuba, Tennodai 1-1-1, Tsukuba, 305-8573, Japan}}
\footnotetext{\textit{$^{c}$~London Centre for Nanotechnology, University College London, Gower Street, London WC1E 6BT, UK}}
\footnotetext{\textit{$^{d}$~Center for Computational Sciences, University of Tsukuba, Tsukuba, 305-8577, Japan}}
\footnotetext{\textit{$^{e}$~JASRI/ SPring-8, 1-1-1 Koto, Sayo-cho, Sayo-gun, Hyogo 679-5198, Japan}}
\footnotetext{\textit{$^{f}$~International Centre for Materials Nanoarchitectonics (MANA), National Institute for Materials Science (NIMS) 1-1 Namiki, Tsukuba, 305-0044, Japan}}
\footnotetext{\textit{$^{g}$~Department of Electrical Materials and Engineering, University of Hyogo, Shosya 2167, Himeji, Hyogo, 671-2280, Japan}}
\footnotetext{\dag~Electronic Supplementary Information (ESI) available: [details of any supplementary information available should be included here]. See DOI: 10.1039/b000000x/}
\footnotetext{\ddag~Contributed equally to this work}
\footnotetext{\textsection~Present address : Central Research Institute of Electric Power Industry, 2-6-1 Nagasaka, Yokosuka, Kanagawa, 240-0196, Japan}


\section{Introduction}
Transition metal doped group IV semiconductors, e.g. Si and Ge, have great potential as dilute magnetic semiconductors (DMS) for spintronic applications with complementary metal-oxide-semiconductor (CMOS) compatibility~\cite{PhysRevB.68.155203, ma3125054, Nie2015279}. Structural engineering is being intensively explored as a method to enhance the Curie temperature (T$_c$) of group IV based DMSs~\cite{Nie2016, C6NR08688H}. However, compared to III-V and II-VI semiconductors, e.g. InAs~\cite{PhysRevLett.68.2664}, the practical realization of a DMS is difficult because of a tendency to form compounds. Typically, transition metals will easily form Si compounds, i.e. silicides~\cite{Allam201369, Lippitz2005307, PhysRevB.88.174419}. For example, Mn$_4$Si$_7$, which forms on the Si(001) surface by Mn deposition at an elevated temperature of around 400$^\circ$C~\cite{0953-8984-24-9-095005}. Thermal processes must be avoided to prevent silicide formation. Ion implantation, an alternative doping method, cannot be used since high temperature annealing is unavoidable. Limitation of germanide formation has been more successful, and Ge based DMSs achieved via techniques such as subsurfactant epitaxy~\cite{Zeng2008}, or Mn doping of quantum dots~\cite{Xiu2010}. However silicides were found to form when combined with Si~\cite{Prestat2014}.

Recently, Mn was found to form atomic chains which run perpendicular to the Si dimer rows on Si(001)~\cite{Liu2008986, jp105620d, PhysRevLett.109.146102}. Their precise physical structure has been a matter of debate, due to their complicated appearance in scanning tunneling microscopy (STM) images~\cite{PhysRevLett.105.116102, PhysRevLett.115.256104}. The current leading model consists of a single Mn atom per chain unit, adsorbed at the edge of the dimer rows, and between two Si dimers. Deposition of sub-monolayer (ML) Mn at a relatively low temperature, i.e. room temperature (RT), leads to these Mn atomic chains on Si(001) without silicide formation. Here, our interest is in using this nanostucture as a Mn dopant source. Theoretical works have predicted that a sub-ML of Mn in the Si lattice, known as a digital magnetic alloy, can be a half metal~\cite{PhysRevLett.96.027211, PhysRevLett.98.117202}. We have previously demonstrated the effectiveness of using surface nanostructures as a dopant source for realizing a highly concentrated Bi $\delta$-doped layer in Si~\cite{Mikidelta, 0953-8984-29-15-155001}.

In this letter, we shall reveal the structure of Mn $\delta$-doped layers based on a combination of X-ray absorption fine structure (XAFS) analysis and density functional theory (DFT) calculations. Using single domain Si(001) surfaces we fabricate aligned Mn atomic chains, which enables us to study anisotropic structures by polarized XAFS techniques. By studying the local structure of samples post-annealing, we demonstrate the thermal stability of the Mn $\delta$-doped layer, and that encapsulation at room temperature ensures that silicide/germanide formation is prevented. We will also demonstrate the magnetic ordering of the $\delta$-doped layer through magnetotransport measurements.

\section{Experimental and Theoretical Methods}
All samples were grown by an ion-pumped Si molecular beam epitaxy (MBE) system with an electron-beam evaporator for Si growth and resistively-heated effusion cells for the evaporation of Mn and Ge. The base pressure was 2$\times$10$^{-7}$~Pa. The growth rate of Si was 15~\AA/min. Mn flux was 3$\times$10$^{13}$ atoms cm$^{-2}$s$^{-1}$. A 20~keV reflection high-energy electron diffraction (RHEED) system was used for monitoring surface structures. A highly resistive Si(001) surface 4$^\circ$ off towards [110] from the (001) substrate was subjected to a standard cleaning process~\cite{Miki1998312}. This results in a single domain Si(001) surface, for the fabrication of single domain Mn atomic chains.

The substrate was immediately loaded into the growth chamber. A Si(001) 2$\times$1 surface was established by growth of a 100~nm thick buffer layer followed by thermal annealing at 800$^\circ$C. The substrate temperature was then ramped down to room temperature and 0.5~ML of Mn was deposited. Here, 1~ML = 6.8 $\times$ 10$^{14}$ atoms cm$^{-2}.$ This dosage was optimized to achieve the highest ratio of Mn chains~\cite{jp105620d}. Finally, a 50~nm-thick Si, or 20~nm-thick Ge and 30~nm-thick Si capping layer was deposited at room temperature. Their interfaces were investigated with transmission electron microscopy (TEM; JEOL JEM 2100F).

XAFS measurements were performed at the Mn K-edge in N$_2$ at the beamline BL37XU, in SPring-8. The X-rays from the undulator source were monochromatized by a Si(111) double-crystal monochromator. The incident X-ray intensity was measured by an ionization chamber. The XAFS spectra were measured in the wavelength dispersive fluorescence mode using a 19-element Ge solid state detector. The sample was mounted on a swivel stage to set the incident angle between the X-rays and the sample surface at 4$^\circ$. The measurement direction was defined by two angles, $\phi$ and $\theta$, between the electric field vector ${\bm E}$ and the Mn atomic chain. $\phi$ is the in-plane angle, such that $\phi = 90^\circ$ when ${\bm E}$ is perpendicular to the Mn chain, and $0^\circ$ when parallel. $\theta$ is the out-of plane is angle, such that $\theta = 0^\circ$ when ${\bm E}$ is perpendicular to the sample normal. The XAFS spectra were analyzed using the Demeter software~\cite{Ravel:ph5155}. Theoretical scattering phase shift and amplitudes were calculated using FEFF8~\cite{PhysRevB.58.7565}. Curve fitting was used to determine the distances (R), effective coordination numbers (CN$^*$), Debye-Waller factors ($\sigma^2$), and relative error between the fit and data (R-factor) for each sample, with an energy shift parameter $\Delta$E used to align the theoretical spectrum to the measured spectrum. Since polarized X-rays were used, the CN$^*$ has a polarization dependence (CN$^*$=$\sum 3N_i\cos^2\theta_i$) for anisotropic specimens. Here $N_i$ is the original coordination number and $\theta_i$ is the angle between the polarization of the X-rays and the bond direction of the target atoms (i.e. Mn) and neighboring atoms. Thus, if $\theta_i = 90$, CN$^*$ is minimized, ideally zero, even if neighboring atoms exist. In other words, studying the polarization dependence can determine the local structure in anisotropic specimens, like those discussed in this work.

For electrical transport measurements, the specimens were processed into Hall-bar structures (70$\times$600~$\mu$m$^2$) by standard micro-fabrication techniques (i.e. photolithography, dry-etching, and metal evaporation). Prior to the fabrication process, the surface oxide layer of the specimen was removed via chemical etching. Since the Mn $\delta$-doped layer is 100~nm beneath the surface, 200~nm deep contact holes were opened by dry etching before the deposition of 200~nm-thick Al films for electrodes. Finally, the specimens were annealed at 400$^\circ$C in Ar + H$_2$(3\%) for 10 minutes. Electrical measurements were carried out by the four-terminal geometry using both a probe station (TKS-MP-V300-20mm; Toei Scientific Industrial Co., Ltd) with Semiconductor Analyzer (B1500A; Agilent Technology) and the physical property measurement system (PPMS; Quantum Design). 

Calculations were performed using density functional theory (DFT)~\cite{Hohenberg1964,Kohn1965} as implemented in the Vienna \textit{Ab initio} Simulation Package (VASP), 5.4.1~\cite{Kresse1993,Kresse1996}. The core electrons were described using the projector augmented wave (PAW) method~\cite{Kresse1999} with the gradient corrected Perdew-Burke-Ernzerhof (PBE) exchange-correlation functional~\cite{Perdew1996}. The following PAWs were used: 05/Jan/2001 Ge, 05/Jan/2001 Si, 15/Jun/2001 H, 06/Sept/2000 Mn. The 4s and 5d electrons of Mn were treated as valence, the rest as core. An energy cut-off of 340~eV was used, with a (3$\times$3$\times$1) Monkhorst-Pack k-point mesh. All calculations were spin polarized, with both anti-ferromagnetic (AFM) and ferromagnetic (FM) arrangements of Mn considered. A calculated Si lattice constant of 5.469~\AA~was used throughout.

\begin{figure}[b]
	\begin{center}
		\includegraphics[width = \linewidth]{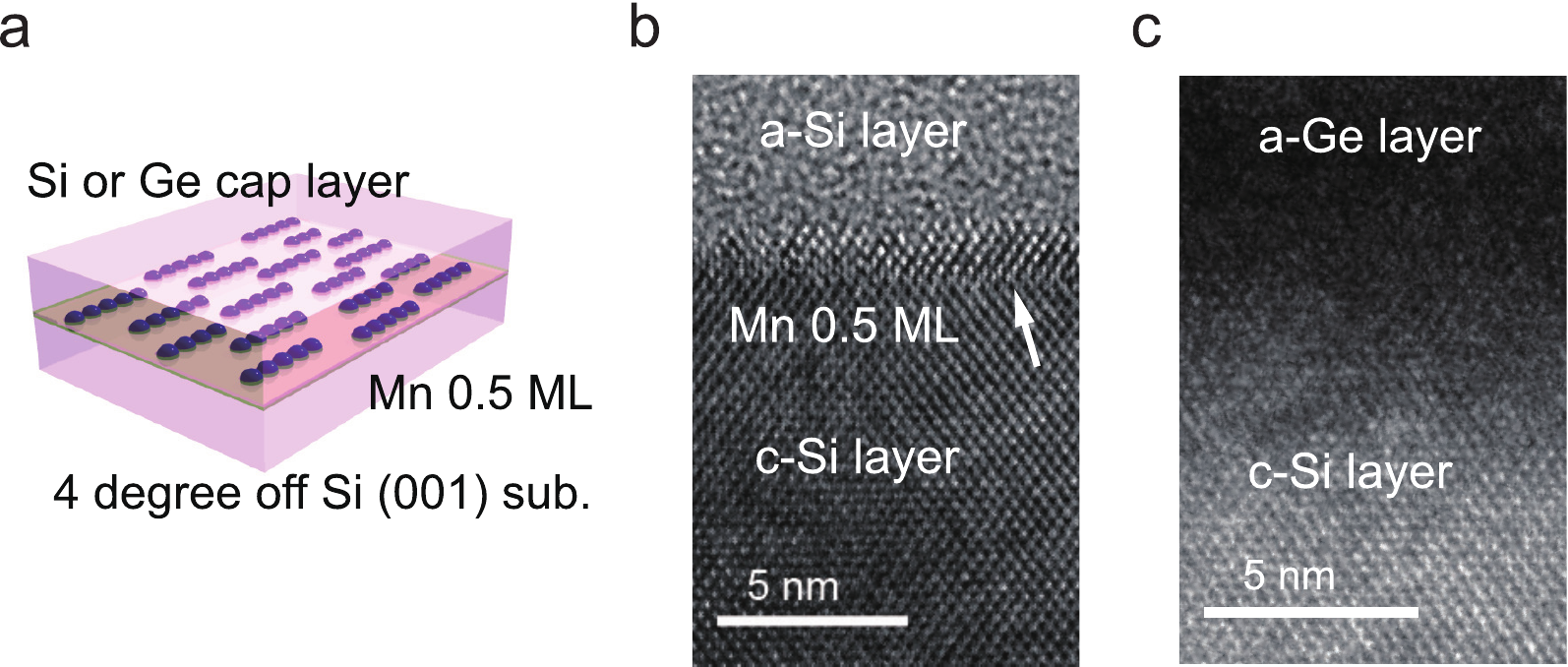}
		\caption{\label{TEM} Structure of the Mn $\delta$-doped layer in the Si crystal and at the Si/Ge interface. a) Schematic model of the sample. Mn atomic chains form in the same direction before burial in a 50~nm-thick Si or 20~nm-thick Ge capping layer. TEM images of samples capped by b) Si and c) Ge layers. Crystalline and amorphous regions are indicated by c-Si and a-Si or a-Ge respectively.}
	\end{center}
\end{figure}

We used a twelve layer slab model, consisting of an eight layer bulk Si substrate and a four layer capping region of Si or Ge, with sixteen atoms per layer. To represent the burial of 0.5~ML of Mn nanowires, we replaced half of the uppermost substrate layer with substitutional Mn, arranged as two lines running parallel to either the $x$ or $y$ axis, separated by intermediate Si atoms. We opted for a slab, rather than bulk, model to avoid creation of a second interfacial region. For Ge capping, we maintained the Si cell parameters, to simulate the effect of the Si substrate. A four layer thick capping region was found to be sufficient to converge the Mn-X (X = Si, Ge) bond lengths to within 0.01~\AA. The bottom layer of the bulk Si region was terminated by H atoms in a dihydride structure, with both Si and H held fixed. The uppermost capping layer was terminated by H in a similar fashion, with the terminating H atoms only allowed to move in $z$. All remaining atoms were allowed to relax until a 0.02~eV/\AA~convergence condition for the forces on each atom was reached. Periodic images were separated by a vacuum gap of 9~\AA~in $z$ to prevent interactions.

\section{Results}
\subsection{Local structure in the $\delta$-doped layer}
TEM images of as-grown samples are shown in Fig.~\ref{TEM} b) and c), for Si and Ge capping respectively. For Si capping, the position of the 0.5~ML Mn $\delta$-doped layer is indicated by an arrow, since several layers of crystalline Si can be grown on Si(001) at T$_{sub} \sim $RT~\cite{Matsuhata}. Whereas for Ge capping, there is a clear distinction between the capping and substrate regions. These images suggest that there are no secondary compounds such as Mn silicides or germanides.

\begin{figure}[tb]
	\begin{center}
		\includegraphics[width = \linewidth]{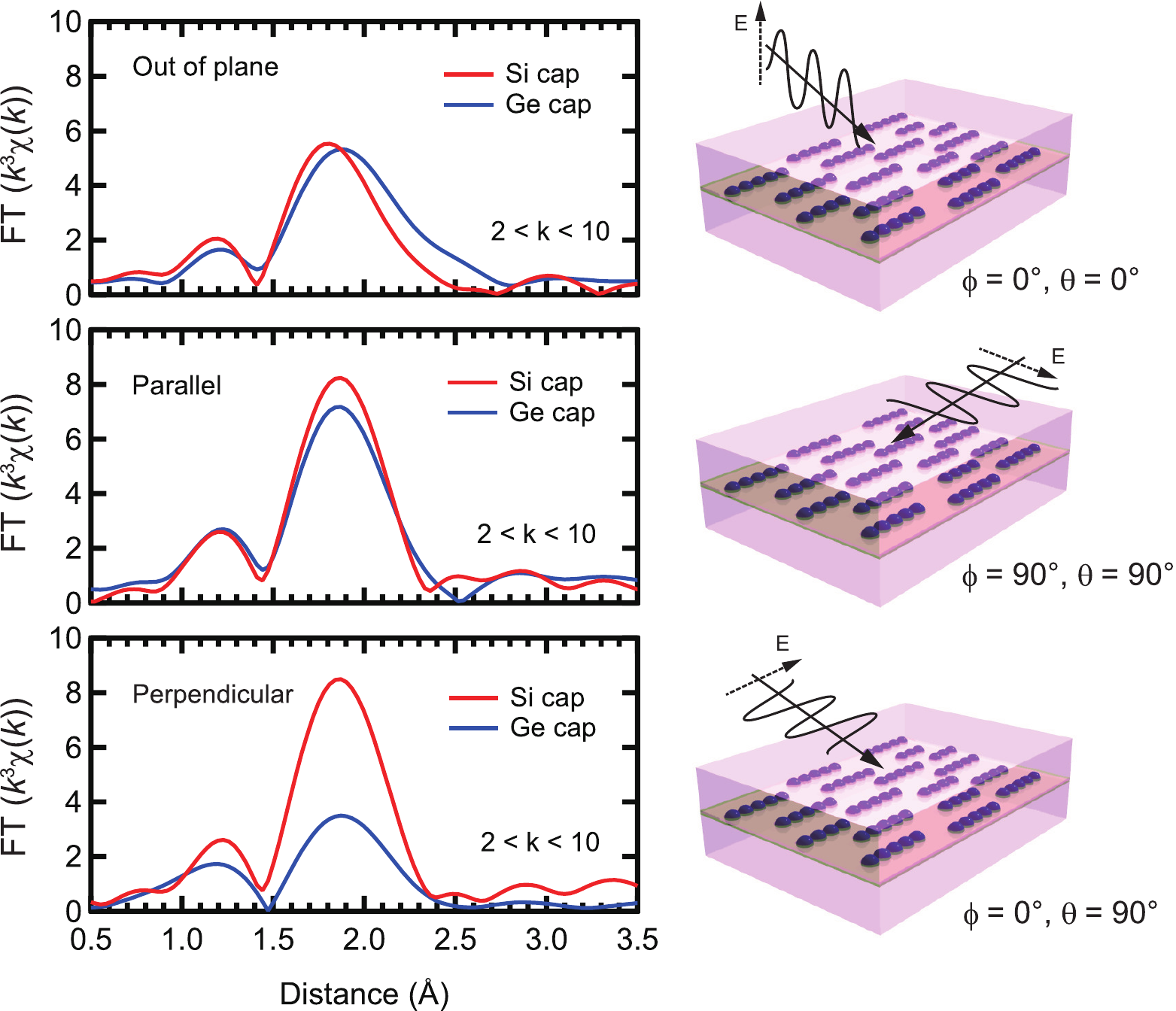}
		\caption{\label{EXAFS_as_grown} Fourier transforms at $k = 2 - 10$~\AA$^{-1}$ of the Mn K-edge EXAFS spectra of Mn $\delta$-doped samples capped by Si and Ge. EXAFS spectra were obtained in three different directions, out-of-plane ($\phi=0^\circ,~\theta=0^\circ$), perpendicular ($\phi=90^\circ,~\theta=90^\circ$) and parallel ($\phi=0^\circ,~\theta=90^\circ$).}
	\end{center}
\end{figure}

\begin{table*}
	\begin{center}
		\caption{\label{t2}Structural parameters for the Mn-Si nearest neighbour shell obtained from curve fitting analysis of the Mn K-edge EXAFS spectra of a Si capped Mn $\delta$-doped layer for $2 < k < 10$ \AA$^{-1}$. The parameters R, $CN^*$ and $\sigma^2$ are the distances, effective coordination number and Debye-Waller factor. $\Delta E$ (eV) and R-factor represent the energy shift to align theoretical spectrum to the measured spectrum and the relative error between the fit and data.}
		\begin{tabular}{lccccc}
			\hline
			Samples & R (\AA) & $CN^*$ & $\sigma^2$ (\AA$^2$) & $\Delta E$ (eV) & R-factor\\
			\hline
			Mn-Si (out-of-plane) & $2.38\pm 0.04$ & $3.9 \pm 1.7$ & $0.012 \pm 0.006$ & -0.13 & 0.00281\\
			Mn-Si (perpendicular) & $2.37\pm 0.08$ & $4.2 \pm 1.4$ & $0.008\pm 0.010$ & -0.24 & 0.00923\\
			Mn-Si (parallel) & $2.37\pm 0.03$ & $4.2 \pm 1.4$ & $0.008 \pm 0.005$ & 0.01 & 0.00239\\
			\hline
		\end{tabular}
	\end{center}
\end{table*}

\begin{table*}
	\begin{center}
		\caption{\label{t3}Structural parameters for the Mn-Si nearest neighbor shell obtained from curve fitting analysis of the Mn K-edge EXAFS spectra of a Ge capped Mn $\delta$-doped layer for $2 < k < 10$ \AA$^{-1}$.  The parameters R, $CN^*$ and $\sigma^2$ are the distances, effective coordination number and Debye-Waller factor. $\Delta E$ (eV) and R-factor represent the energy shift to align theoretical spectrum to the measured spectrum and the relative error between the fit and data. Mn-Si oscillation is dominant and $CN^*$ is dependent on the measurement geometry, which is based on structural anisotropy.}
		\begin{tabular}{lccccc}
			\hline
			Samples & R (\AA) & $CN^*$ & $\sigma^2$ (\AA$^2$) & $\Delta E$ (eV) & R-factor\\
			\hline
			Mn-Si (out-of-plane)			 & $2.37\pm 0.06$ & $1.3 \pm 1.2$ & $-0.001 \pm 0.009$ & 1.4 & 0.00796\\
			Mn-Si (perpendicular) & $2.37\pm 0.03$ & $2.0 \pm 1.0$ & $0.002\pm 0.005$ & 1.1 & 0.00224\\
			Mn-Si (parallel) & $2.38\pm 0.07$ & $1.3 \pm 1.3$ & $0.005 \pm 0.010$ & 2.4 & 0.00813\\
			\hline
		\end{tabular}
	\end{center}
\end{table*}

Figure~\ref{EXAFS_as_grown} shows Fourier transforms of the Mn K-edge extended XAFS (EXAFS) spectra in three different directions, out-of-plane, in-plane parallel and in-plane perpendicular, with the results of curve fitting listed in Tables~\ref{t2} and~\ref{t3}. The obtained X-ray absorption near-edge (XANES) spectra are shown in the Supporting Information and are similar to previous results~\cite{Wolska2007, Ye_co}. Curve fitting was done using a one-shell model with Mn-Si and/or Mn-Ge scattering paths, or using a two-shell model with Mn-Si and Mn-Ge scattering paths. It should be noted that the $x$-axis in Fig.~\ref{EXAFS_as_grown} does not show the correct distance $R$, because the Fourier transform did not account for the effect of phase shift in the photo-electron scattering process. Thus, the extracted values from curve fitting in Table~\ref{t2} are correct. Since a single domain Si(001) surface was used, the Mn chains all align in the same direction. Comparing the results for Si and Ge capping, there is a clear difference in the perpendicular direction, and minor differences in the other two. Typical Mn silicides, such as MnSi and Mn$_4$Si$_7$, display secondary peaks, which are absent from this data. These results confirm the suggestion from earlier TEM observations, that the Mn atoms form $\delta$-doped layers, not Mn silicides. EXAFS data for typical silicides can be found in the Supporting Information.

In the Si capped sample the Mn-Si bond lengths measured in-plane agree, at 2.37~\AA, regardless of direction, with out-of-plane measurements only 0.01~\AA~different. The effective coordination number CN$^*$ is approximately 4 regardless of measurement direction. Given the associated error bars on each measurement, we assume that the local Mn structure is symmetric in-plane, and near equivalent to out-of-plane. Based on the polarization dependence, the original coordination number $N$ is expected to be 2. 

\begin{figure*}
	\begin{center}
		\includegraphics[width = 0.5\linewidth]{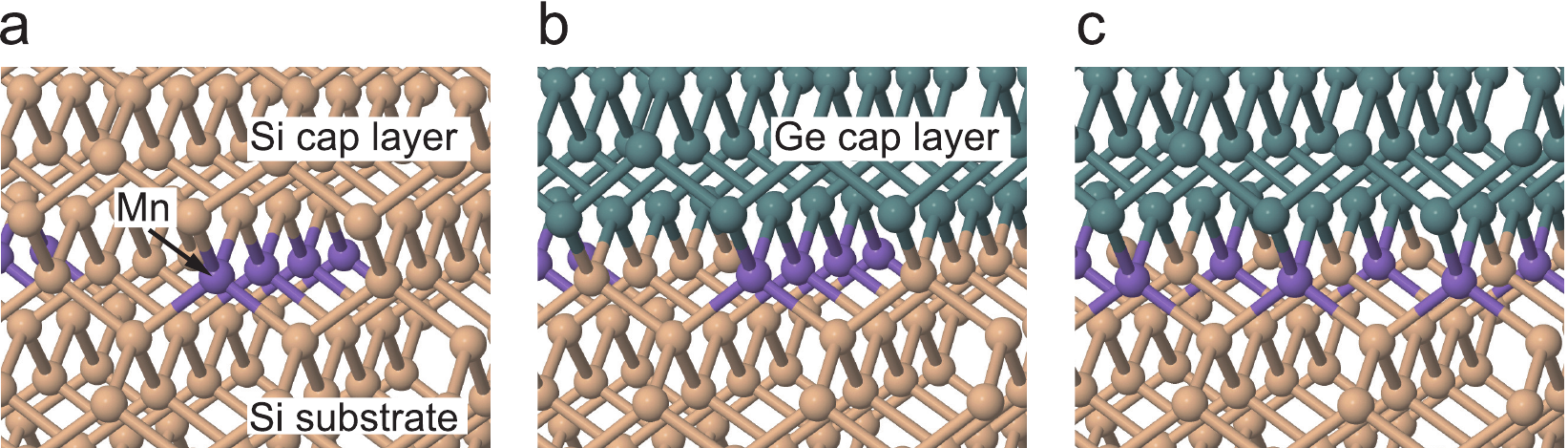}
		\caption{\label{DFT_model}Structural models of the Mn $\delta$-doped layer in a) Si and b), c) the Si/Ge interface. Substitutional Mn atoms form single lines perpendicular to [110] in a) and b), and perpendicular to [100] in c). Optimized geometries for magnetic ordering are shown. Si atoms are in beige, Ge in green and Mn in purple.}
	\end{center}
\end{figure*}

The Ge capped sample introduces complications to the fitting process, because EXAFS cannot distinguish the two elements at a close distance. However, interestingly, the peak height of the radial structure function (RSF) is double in the perpendicular direction compared to the parallel direction. In the perpendicular case, since the RSF is near-identical for Si and Ge capping, we assume that this peak can be attributed to Mn-Si bonding. This means that the RSFs in the perpendicular and parallel directions reflect the local structure below and above the Mn layer, respectively. Below the Mn layer the Mn atoms bond with Si atoms perpendicular to the Mn chain direction, and above the Mn layer they form bonds with Ge atoms parallel to the chain direction. Mn-Si lengths are approximately the same as for Si capping, with the reduction in CN$^*$ to 2 likely due to the formation of Mn-Ge bonds. Furthermore, since $\sigma^2$ is a squared value, the negative value for the out of plane measurement indicates that the result is not reasonable. It should be also be noted that there is a tail at 2.5~\AA~in the RSF for the out of plane measurement, which is discussed below. 

The Mn-Ge parameters had much larger associated error bars, especially for the R-factor and CN$^*$, making it difficult to gain meaningful insight from this data. As such the results are listed in the Supporting Information, and not discussed here.

\begin{figure}
	\begin{center}
		\includegraphics[width=0.5\linewidth]{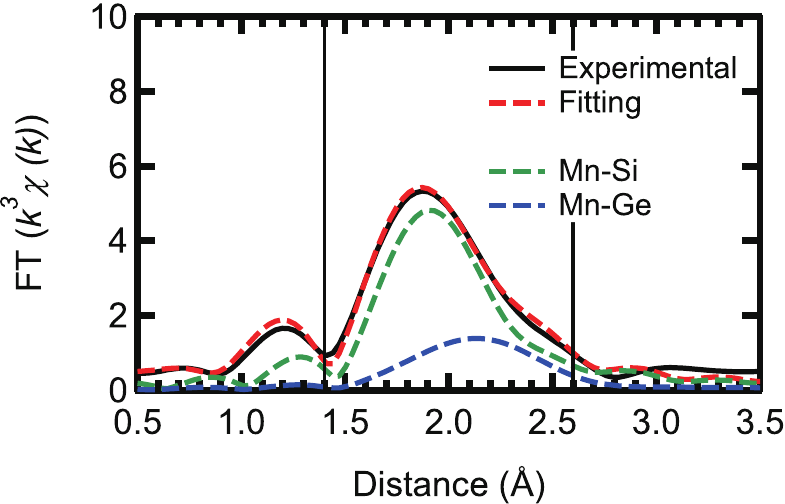}
		\caption{\label{Fitting_two_shell} Curve fitting of the $k^3$-weighted EXAFS spectra (out-of-plane) of a Mn $\delta$ doped layer in the Si/Ge interface. Its Fourier transform is at $k$ = 2 - 10~\AA$^{-1}$, with its curves fitted by a two-shell model.}
	\end{center}
\end{figure}

\begin{table}[htb]
    \begin{center}
	\caption{\label{t2shell}Structural parameters for the Mn-Si and the Mn-Ge nearest neighbour shells obtained from curve fitting analysis of the Mn K-edge EXAFS spectra of a Ge capped Mn $\delta$-doped layer for $2 < k < 10$ \AA$^{-1}$ by a two-shell model. The parameters R, $CN^*$ and $\sigma^2$ are the distances, effective coordination number and Debye-Waller factor.}
	\begin{tabular}{lccc}
		\hline
		Shell & R (\AA) & $CN^*$ & $\sigma^2$ (\AA$^2$) \\
		\hline
		Mn-Si (out-of-plane)			 & $2.37\pm 0.02$ & $1.4 \pm 1.3$ & $0.001 \pm 0.01$\\
		Mn-Ge (out-of-plane) & $2.49\pm 0.03$ & $1.1 \pm 6.3$ & $0.010\pm 0.06$\\
		\hline
	\end{tabular}
	\end{center}
\end{table}

To gain insights into the local structure around the Mn atoms, we calculated Mn-Si/Ge bond lengths for an Si capped system and two variations of Ge capping, depending on the direction of the Mn-Ge bonds, as shown in Fig.~\ref{DFT_model}. For Si capping, the Mn-Si bond length was calculated to be 2.39$\pm$0.01~\AA~with minor variations depending on the direction of the Mn-Si bonds relative to the Mn chains. Bonds parallel to the Mn chain were found to be marginally shorter (0.02~\AA), because adjacent Mn are closer in this direction. Our calculated bond lengths show excellent agreement with the EXAFS results, with a discrepancy smaller than the experimental error bars. This suggests that our $\delta$-doped layer contains Mn atoms at substitutional sites within the Si crystal.

For Ge-capping we considered two possible capping situations, one where the Mn-Ge bonds are parallel to the Mn chain, as in b), and one where the Mn-Ge bonds are perpendicular to the Mn chain, as in c). Mn-Si bond lengths were unaffected by Ge capping. The Mn-Ge bonds are slightly longer than Mn-Si, at 2.435$\pm$0.015~\AA~depending on direction. As for Si capping, these results support the idea that the Mn $\delta$-doped layer contains Mn atoms at substitutional sites in the Si/Ge interface. 

Based on our theoretical model, we performed further fitting, using a two shell model, for the out of plane measurement of the Ge capped sample. This fitting, for both Mn-Si and Mn-Ge, is shown in Fig.~\ref{Fitting_two_shell} with the obtained parameters listed in Table~\ref{t2shell}. To reduce the number of fitting parameters, we initially fixed the Mn-Si and Mn-Ge bond lengths. Then using the obtained CN$^*$ and $\sigma^2$ values, bond lengths were fitted. Unfortunately, the CN$^*$ for Mn-Ge still has large error bars. However, $\sigma^2$ is now positive and the R-factor improved to 0.00531. Thus the results of the present fitting shows good agreement with both the experimental and theoretical aspects, although there is still some arbitrariness to how the fitting is performed.

Based on the measured and calculated structural parameters, we have shown that a Mn $\delta$-doped layer, with Mn atoms at substitutional sites, has been achieved, without silicide formation in both capping environments. It should be emphasized that this structure cannot be achieved with other doping techniques.

\subsection{Thermal stability: annealing effects}
The thermal stability, or instability, of the $\delta$-doped layer is an important issue when investigating its electrical and magnetic properties, since transition metals are known to easily form compounds such as silicides and germanides as described above. Attempts to make a DMS via ion implantation cannot avoid high temperature annealing, which causes most of the Mn ions to form a silicide.

\begin{figure}
	\begin{center}
		\includegraphics[width = 0.6\linewidth]{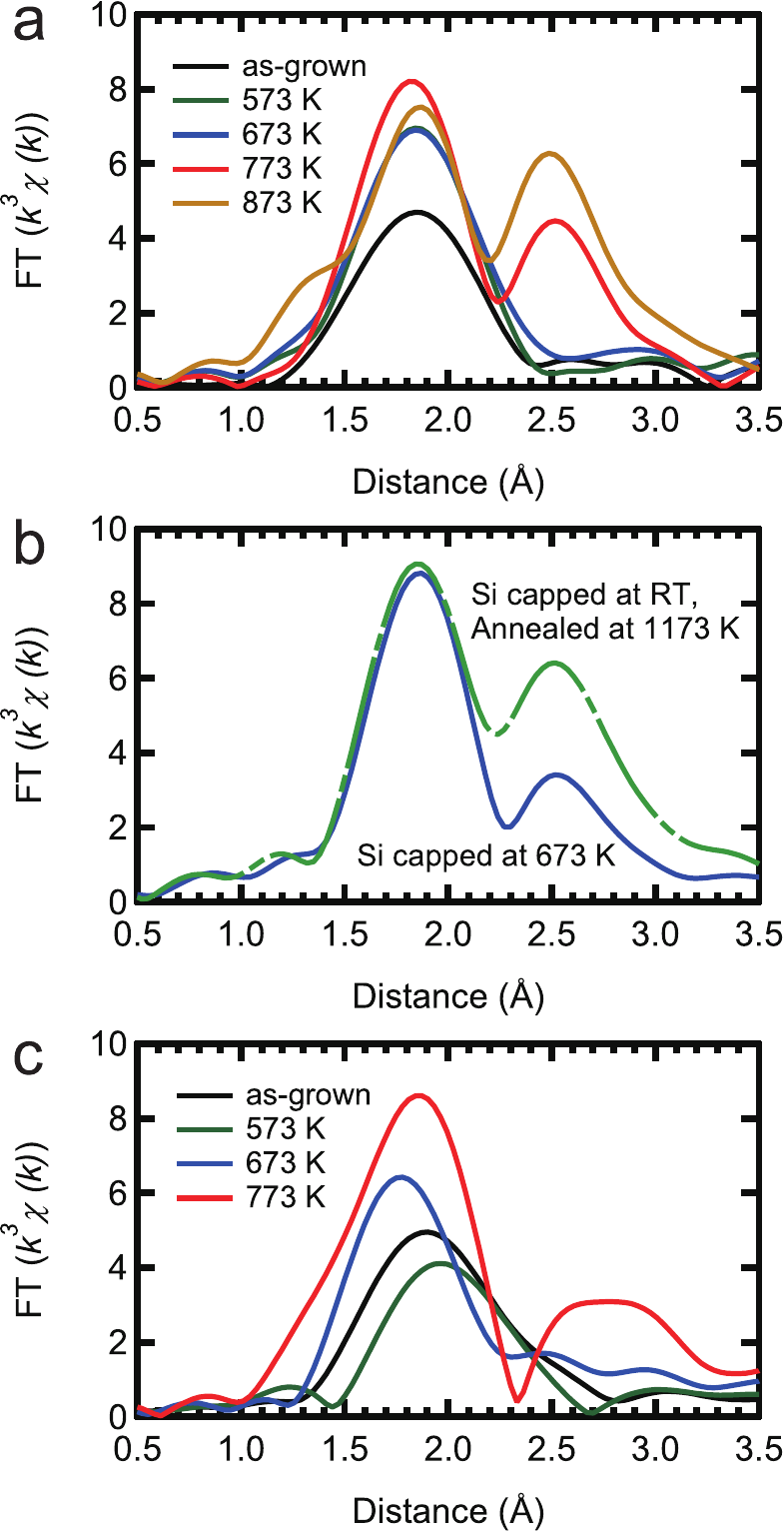}
		\caption{\label{silicidation} Annealing effect on the Mn $\delta$-doped layer. Fourier transform at $k = 2 - 10$ \AA$^{-1}$ of the Mn K-edge EXAFS spectra of samples capped with a) Si and c) Ge. b) Reference data for a sample capped with Si at 673~K and annealed at a higher temperature (1173~K). Plot of the peak intensity of the second nearest neighbor shell. Both samples show clear evidence for the formation of Mn silicide after annealing above 673~K for 30 minutes.}
	\end{center}
\end{figure}

The radial structure functions (RSFs) for our samples after annealing at various temperatures are shown in Fig.~\ref{silicidation}. After annealing the Si capped sample at over 773~K, a secondary peak appears at around 2.5~\AA~in the RSF and the peak intensity of the second nearest neighbour shell rises, indicating the formation of a silicide. As a reference for silicidation, RSFs of a sample capped with Si at 673~K and annealed at a higher temperature (1173~K) are shown in Fig.~\ref{silicidation} (b). Here it is clear that the Si capping process at 673~K cannot prevent Mn silicide formation. With respect to silicide formation, our Si capped Mn $\delta$-doped layer is thermally stable up to 673~K. Although, there are visible thermal effect even at 573~K. The Ge capped sample is slightly more complicated. Annealing at higher temperatures can lead to silicide formation, as evidenced by the secondary peak, but it also leads to diffusion of Mn into the Ge layer. Setting aside the complications of high temperature anneals, we can see that our Ge capped samples are also stable up to 673 K.

\subsection{Magnetotransport}
\begin{figure}
	\begin{center}
		\includegraphics[width = \linewidth]{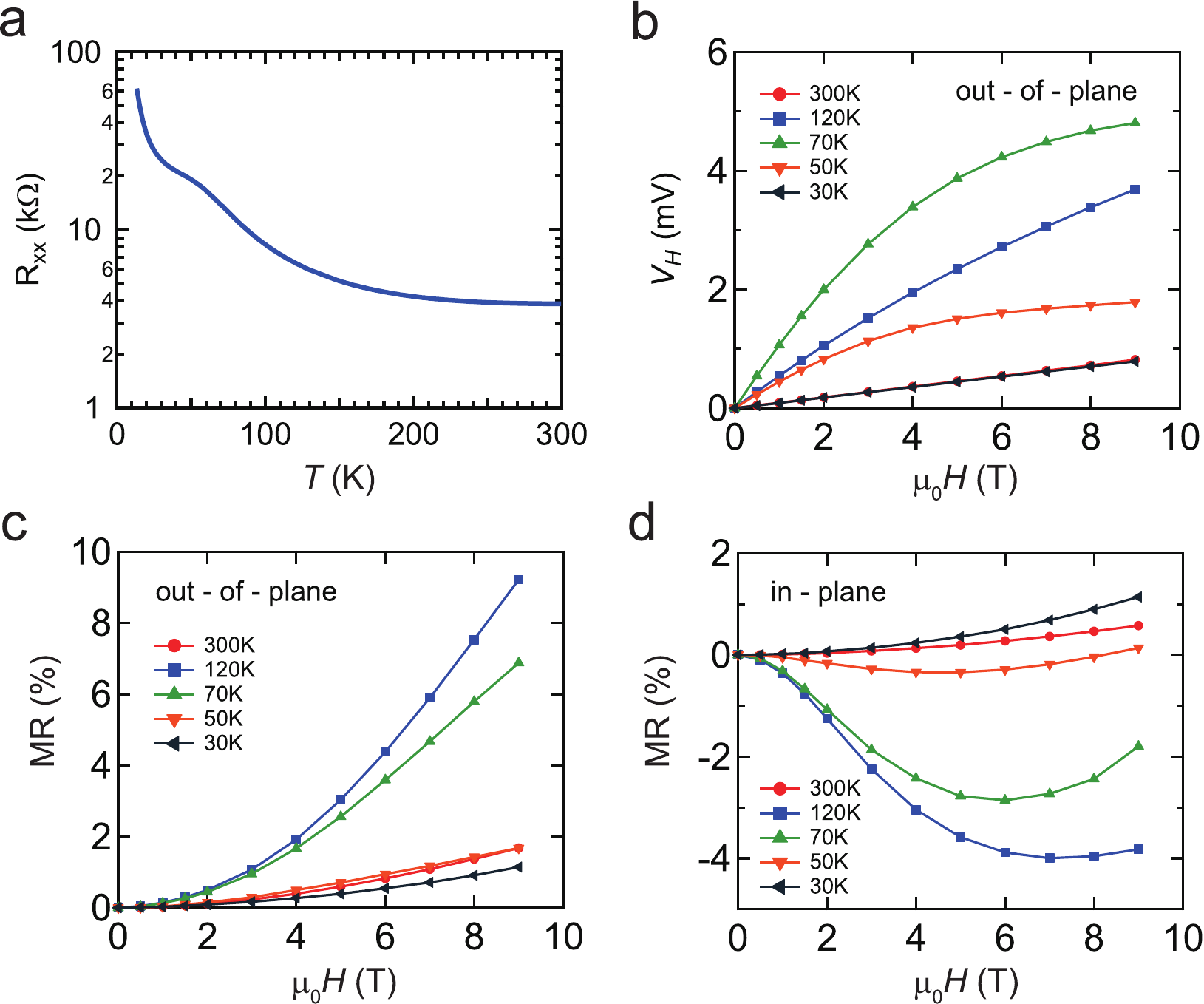}
		\caption{\label{transport}Transport properties of a Mn $\delta$-doped layer capped by a-Si. a) Temperature dependence of magnetoresistance $R_{xx}$, b) external field dependence of Hall voltage, and magnetoresistance $R_{xx}$ measured c) out-of-plane and d) in-plane.}
	\end{center}
\end{figure}

Finally we briefly discuss the transport properties of the Mn $\delta$-doped layer capped by Si, as shown in Fig.~\ref{transport}. Due to difficulties in making ideal ohmic contacts for the Ge capped Mn $\delta$-doped layer, which is sandwiched between Si and Ge, clear transport data could not be obtained at this time. Even for Si capping, the doped layer is atomically thin, i.e. < 1~nm, so transport data will include the properties of both capping and buffer layers. However, the features of the Mn $\delta$-doped layer should dominate the transport properties because of the additional highly concentrated impurities that provide extrinsic carriers. Furthermore, the magnetic response must be from the Mn impurities in the Mn $\delta$-doped layer. 

The temperature dependence of the longitudinal magnetoresistance $R_{xx}$ ($\Omega$) shows semiconductor behavior with an anomaly at around 100~K (Fig.~\ref{transport} a)). The magnetic field, $\mu_0H$(T), dependence of the Hall voltage, $V_H$, demonstrates that the majority carriers are p-type (Fig.~\ref{transport} b)). In addition to an ordinary linear Hall effect, the anomalous Hall effect (AHE) was observed at temperatures of 50 and 70~K. The Hall resistance with AHE component is empirically described as $R_{H}= R_0\mu_0H+R_sM_z$~\cite{RevModPhys.82.1539}, where $M_z$ is the magnetization averaged over the sample, $\mu_0$ is the permeability of vacuum, and $R_0$ and $R_s$ are the ordinary and anomalous Hall coefficients, respectively. Based on the ordinary Hall coefficient $R_0$, we determined that the sheet (2D) carrier density was $6\times10^{13}$cm$^{-2}$ at 300 K. By considering the Mn impurity levels in the Si band gap, that are relatively deep, this carrier density would be due to unintentional B impurities, a common phenomenon in Si MBE epitaxy\cite{Miki1998312}. 

A relatively large positive magnetoresistance (MR) was observed out-of-plane, with a negative MR in-plane. It is well known that negative MR is enhanced near the $T_c$ in metal systems due to spin-disorder scattering. Taken together, these results mean that a magnetic system is observed at up to 120~K, which is in good agreement with theoretical predictions for substitutional Mn in the Si crystal~\cite{PhysRevB.63.195205}. This is supported by DFT calculations for our Mn $\delta$-doped systems, which showed that magnetic ordering was energetically favored. Our observations cannot be explained by silicide formation, since the $T_c$ of MnSi and Mn$_4$Si$_7$ are 29 and 40~K respectively. The origin of the magnetism is expected to be the Hole-mediated Ruderman-Kittel-Kasuya-Yosida (RKKY) interaction. The distance between Mn atoms is large enough that localized spins will not interact with each other. Moreover, there is a peak in the temperature dependence of the magnetic response. This temperature dependence could be explained by considering the interplay between two factors, (1) the suppression of thermal fluctuation which enhance the magnetic order and (2) the decrease of carrier density which weakens the RKKY interaction. Therefore, co-doping with more acceptor atoms or applying a field effect are important for verifying this hypothesis.

\section{Conclusion}
We demonstrate successful $\delta$-doping of Mn into Si and a Si/Ge interface based on capping of Mn atomic chains on Si(001). Silicide / germanide formation is avoided by encapsulation at room temperature. XAFS techniques combined with DFT calculations reveal that our samples contain Mn atoms at substitutional sites within the Si/Ge lattice. A comprehensive magnetotransport study reveals a magnetic ordering within the Mn $\delta$-doped layer. Doping methods based on the burial of surface nanostructures paves the way towards realizing the ultimate doping profile for transition metals; atomic scale doping in the desired local state, which is impossible with conventional methods.

\section*{Acknowledgement}
The synchrotron radiation experiments were performed at the beamline BL37XU, SPring-8 with the approval of the Japan Synchrotron Radiation Research Institute (JASRI) (Proposal Nos.2013A1274 and 2014B1097). Calculations were performed using the UCL Legion High Performance Computing Facility, and associated support services. This research was supported by JSPS KAKENHI Grant Numbers 20246017, 17H05225, 17H02777 (K.Miki) and 20121355 (K. Murata), and also NIMS Nanofabrication Platform in Nanotechnology Platform Project from the Ministry of Education, Culture, Sports, Science and Technology, Japan (MEXT).  C. J. Kirkham was funded through a UCL Impact Studentship.




\bibliography{Paper_Mn_doping_Si} 

\providecommand*{\mcitethebibliography}{\thebibliography}
\csname @ifundefined\endcsname{endmcitethebibliography}
{\let\endmcitethebibliography\endthebibliography}{}
\begin{mcitethebibliography}{36}
\providecommand*{\natexlab}[1]{#1}
\providecommand*{\mciteSetBstSublistMode}[1]{}
\providecommand*{\mciteSetBstMaxWidthForm}[2]{}
\providecommand*{\mciteBstWouldAddEndPuncttrue}
  {\def\EndOfBibitem{\unskip.}}
\providecommand*{\mciteBstWouldAddEndPunctfalse}
  {\let\EndOfBibitem\relax}
\providecommand*{\mciteSetBstMidEndSepPunct}[3]{}
\providecommand*{\mciteSetBstSublistLabelBeginEnd}[3]{}
\providecommand*{\EndOfBibitem}{}
\mciteSetBstSublistMode{f}
\mciteSetBstMaxWidthForm{subitem}
{(\emph{\alph{mcitesubitemcount}})}
\mciteSetBstSublistLabelBeginEnd{\mcitemaxwidthsubitemform\space}
{\relax}{\relax}

\bibitem[Stroppa \emph{et~al.}(2003)Stroppa, Picozzi, Continenza, and
  Freeman]{PhysRevB.68.155203}
A.~Stroppa, S.~Picozzi, A.~Continenza and A.~J. Freeman, \emph{Phys. Rev. B},
  2003, \textbf{68}, 1552031--1552039\relax
\mciteBstWouldAddEndPuncttrue
\mciteSetBstMidEndSepPunct{\mcitedefaultmidpunct}
{\mcitedefaultendpunct}{\mcitedefaultseppunct}\relax
\EndOfBibitem
\bibitem[Zhou and Schmidt(2010)]{ma3125054}
S.~Zhou and H.~Schmidt, \emph{Materials}, 2010, \textbf{3}, 5054--5082\relax
\mciteBstWouldAddEndPuncttrue
\mciteSetBstMidEndSepPunct{\mcitedefaultmidpunct}
{\mcitedefaultendpunct}{\mcitedefaultseppunct}\relax
\EndOfBibitem
\bibitem[Nie \emph{et~al.}(2015)Nie, Tang, and Wang]{Nie2015279}
T.~Nie, J.~Tang and K.~L. Wang, \emph{Journal of Crystal Growth}, 2015,
  \textbf{425}, 279--282\relax
\mciteBstWouldAddEndPuncttrue
\mciteSetBstMidEndSepPunct{\mcitedefaultmidpunct}
{\mcitedefaultendpunct}{\mcitedefaultseppunct}\relax
\EndOfBibitem
\bibitem[Nie \emph{et~al.}(2016)Nie, Tang, Kou, Gen, Lee, Zhu, He, Chang,
  Murata, Fan, and Wang]{Nie2016}
T.~Nie, J.~Tang, X.~Kou, Y.~Gen, S.~Lee, X.~Zhu, Q.~He, L.-T. Chang, K.~Murata,
  Y.~Fan and K.~L. Wang, \emph{Nature Communications}, 2016, \textbf{7},
  12866\relax
\mciteBstWouldAddEndPuncttrue
\mciteSetBstMidEndSepPunct{\mcitedefaultmidpunct}
{\mcitedefaultendpunct}{\mcitedefaultseppunct}\relax
\EndOfBibitem
\bibitem[Nie \emph{et~al.}(2017)Nie, Kou, Tang, Fan, Lee, He, Chang, Murata,
  Gen, and Wang]{C6NR08688H}
T.~Nie, X.~Kou, J.~Tang, Y.~Fan, S.~Lee, Q.~He, L.-T. Chang, K.~Murata, Y.~Gen
  and K.~L. Wang, \emph{Nanoscale}, 2017, \textbf{9}, 3086--3094\relax
\mciteBstWouldAddEndPuncttrue
\mciteSetBstMidEndSepPunct{\mcitedefaultmidpunct}
{\mcitedefaultendpunct}{\mcitedefaultseppunct}\relax
\EndOfBibitem
\bibitem[Ohno \emph{et~al.}(1992)Ohno, Munekata, Penney, von Moln\'ar, and
  Chang]{PhysRevLett.68.2664}
H.~Ohno, H.~Munekata, T.~Penney, S.~von Moln\'ar and L.~L. Chang, \emph{Phys.
  Rev. Lett.}, 1992, \textbf{68}, 2664--2667\relax
\mciteBstWouldAddEndPuncttrue
\mciteSetBstMidEndSepPunct{\mcitedefaultmidpunct}
{\mcitedefaultendpunct}{\mcitedefaultseppunct}\relax
\EndOfBibitem
\bibitem[Allam \emph{et~al.}(2013)Allam, Boulet, and Record]{Allam201369}
A.~Allam, P.~Boulet and M.-C. Record, \emph{Journal of Electronic Materials},
  2013, \textbf{43}, 761--773\relax
\mciteBstWouldAddEndPuncttrue
\mciteSetBstMidEndSepPunct{\mcitedefaultmidpunct}
{\mcitedefaultendpunct}{\mcitedefaultseppunct}\relax
\EndOfBibitem
\bibitem[Lippitz \emph{et~al.}(2005)Lippitz, Paggel, and
  Fumagalli]{Lippitz2005307}
H.~Lippitz, J.~Paggel and P.~Fumagalli, \emph{Surf. Sci.}, 2005, \textbf{575},
  307--312\relax
\mciteBstWouldAddEndPuncttrue
\mciteSetBstMidEndSepPunct{\mcitedefaultmidpunct}
{\mcitedefaultendpunct}{\mcitedefaultseppunct}\relax
\EndOfBibitem
\bibitem[Kahwaji \emph{et~al.}(2013)Kahwaji, Gordon, Crozier, Roorda,
  Robertson, Zhu, and Monchesky]{PhysRevB.88.174419}
S.~Kahwaji, R.~A. Gordon, E.~D. Crozier, S.~Roorda, M.~D. Robertson, J.~Zhu and
  T.~L. Monchesky, \emph{Phys. Rev. B}, 2013, \textbf{88}, 174419\relax
\mciteBstWouldAddEndPuncttrue
\mciteSetBstMidEndSepPunct{\mcitedefaultmidpunct}
{\mcitedefaultendpunct}{\mcitedefaultseppunct}\relax
\EndOfBibitem
\bibitem[Liu \emph{et~al.}(2012)Liu, Owen, and Miki]{0953-8984-24-9-095005}
H.~J. Liu, J.~H.~G. Owen and K.~Miki, \emph{J. Phys.: Condens. Matter}, 2012,
  \textbf{24}, 095005\relax
\mciteBstWouldAddEndPuncttrue
\mciteSetBstMidEndSepPunct{\mcitedefaultmidpunct}
{\mcitedefaultendpunct}{\mcitedefaultseppunct}\relax
\EndOfBibitem
\bibitem[Zeng \emph{et~al.}(2008)Zeng, Zhang, van Benthem, Chisholm, and
  Weitering]{Zeng2008}
C.~Zeng, Z.~Zhang, K.~van Benthem, M.~F. Chisholm and H.~H. Weitering,
  \emph{Phys. Rev. Lett.}, 2008, \textbf{100}, 066101\relax
\mciteBstWouldAddEndPuncttrue
\mciteSetBstMidEndSepPunct{\mcitedefaultmidpunct}
{\mcitedefaultendpunct}{\mcitedefaultseppunct}\relax
\EndOfBibitem
\bibitem[Xiu \emph{et~al.}(2010)Xiu, Wang, Kim, Hong, Tang, Jacob, Zou, and
  Wang]{Xiu2010}
F.~Xiu, Y.~Wang, J.~Kim, A.~Hong, J.~Tang, A.~P. Jacob, J.~Zou and K.~L. Wang,
  \emph{Nat. Mater.}, 2010, \textbf{9}, 337\relax
\mciteBstWouldAddEndPuncttrue
\mciteSetBstMidEndSepPunct{\mcitedefaultmidpunct}
{\mcitedefaultendpunct}{\mcitedefaultseppunct}\relax
\EndOfBibitem
\bibitem[Prestat \emph{et~al.}(2014)Prestat, Porret, Favre-Nicolin, Tainoff,
  Boukhari, Bayle-Guillemaud, Jamet, and Barski]{Prestat2014}
E.~Prestat, C.~Porret, V.~Favre-Nicolin, D.~Tainoff, M.~Boukhari,
  P.~Bayle-Guillemaud, M.~Jamet and A.~Barski, \emph{Appl. Phys. Lett.}, 2014,
  \textbf{104}, 102409\relax
\mciteBstWouldAddEndPuncttrue
\mciteSetBstMidEndSepPunct{\mcitedefaultmidpunct}
{\mcitedefaultendpunct}{\mcitedefaultseppunct}\relax
\EndOfBibitem
\bibitem[Liu and Reinke(2008)]{Liu2008986}
H.~Liu and P.~Reinke, \emph{Surf. Sci.}, 2008, \textbf{602}, 986--992\relax
\mciteBstWouldAddEndPuncttrue
\mciteSetBstMidEndSepPunct{\mcitedefaultmidpunct}
{\mcitedefaultendpunct}{\mcitedefaultseppunct}\relax
\EndOfBibitem
\bibitem[Nolph \emph{et~al.}(2010)Nolph, Simov, Liu, and Reinke]{jp105620d}
C.~A. Nolph, K.~R. Simov, H.~Liu and P.~Reinke, \emph{J. Phys. Chem. C}, 2010,
  \textbf{114}, 19727--19733\relax
\mciteBstWouldAddEndPuncttrue
\mciteSetBstMidEndSepPunct{\mcitedefaultmidpunct}
{\mcitedefaultendpunct}{\mcitedefaultseppunct}\relax
\EndOfBibitem
\bibitem[Fuhrer \emph{et~al.}(2012)Fuhrer, Rue\ss{}, Moll, Curioni, and
  Widmer]{PhysRevLett.109.146102}
A.~Fuhrer, F.~J. Rue\ss{}, N.~Moll, A.~Curioni and D.~Widmer, \emph{Phys. Rev.
  Lett.}, 2012, \textbf{109}, 146102\relax
\mciteBstWouldAddEndPuncttrue
\mciteSetBstMidEndSepPunct{\mcitedefaultmidpunct}
{\mcitedefaultendpunct}{\mcitedefaultseppunct}\relax
\EndOfBibitem
\bibitem[Wang \emph{et~al.}(2010)Wang, Chen, Wang, and
  Kawazoe]{PhysRevLett.105.116102}
J.-T. Wang, C.~Chen, E.~Wang and Y.~Kawazoe, \emph{Phys. Rev. Lett.}, 2010,
  \textbf{105}, 116102\relax
\mciteBstWouldAddEndPuncttrue
\mciteSetBstMidEndSepPunct{\mcitedefaultmidpunct}
{\mcitedefaultendpunct}{\mcitedefaultseppunct}\relax
\EndOfBibitem
\bibitem[Villarreal \emph{et~al.}(2015)Villarreal, Longobardi, K\"oster,
  Kirkham, Bowler, and Renner]{PhysRevLett.115.256104}
R.~Villarreal, M.~Longobardi, S.~A. K\"oster, C.~J. Kirkham, D.~Bowler and
  C.~Renner, \emph{Phys. Rev. Lett.}, 2015, \textbf{115}, 256104\relax
\mciteBstWouldAddEndPuncttrue
\mciteSetBstMidEndSepPunct{\mcitedefaultmidpunct}
{\mcitedefaultendpunct}{\mcitedefaultseppunct}\relax
\EndOfBibitem
\bibitem[Qian \emph{et~al.}(2006)Qian, Fong, Liu, Pickett, Pask, and
  Yang]{PhysRevLett.96.027211}
M.~C. Qian, C.~Y. Fong, K.~Liu, W.~E. Pickett, J.~E. Pask and L.~H. Yang,
  \emph{Phys. Rev. Lett.}, 2006, \textbf{96}, 027211\relax
\mciteBstWouldAddEndPuncttrue
\mciteSetBstMidEndSepPunct{\mcitedefaultmidpunct}
{\mcitedefaultendpunct}{\mcitedefaultseppunct}\relax
\EndOfBibitem
\bibitem[Wu \emph{et~al.}(2007)Wu, Kratzer, and
  Scheffler]{PhysRevLett.98.117202}
H.~Wu, P.~Kratzer and M.~Scheffler, \emph{Phys. Rev. Lett.}, 2007, \textbf{98},
  117202\relax
\mciteBstWouldAddEndPuncttrue
\mciteSetBstMidEndSepPunct{\mcitedefaultmidpunct}
{\mcitedefaultendpunct}{\mcitedefaultseppunct}\relax
\EndOfBibitem
\bibitem[Miki \emph{et~al.}(1999)Miki, Matsuhata, Sakamoto, Briggs, Owen, and
  Bowler]{Mikidelta}
K.~Miki, H.~Matsuhata, K.~Sakamoto, G.~A.~D. Briggs, J.~H.~G. Owen and D.~R.
  Bowler, \emph{Inst. Phys. Conf. Ser.}, 1999, \textbf{164}, 167\relax
\mciteBstWouldAddEndPuncttrue
\mciteSetBstMidEndSepPunct{\mcitedefaultmidpunct}
{\mcitedefaultendpunct}{\mcitedefaultseppunct}\relax
\EndOfBibitem
\bibitem[Murata \emph{et~al.}(2017)Murata, Kirkham, Shimomura, Nitta, Uruga,
  Terada, ichi Nittoh, Bowler, and Miki]{0953-8984-29-15-155001}
K.~Murata, C.~Kirkham, M.~Shimomura, K.~Nitta, T.~Uruga, Y.~Terada, K.~ichi
  Nittoh, D.~R. Bowler and K.~Miki, \emph{J. Phys.: Condens. Matter}, 2017,
  \textbf{29}, 155001\relax
\mciteBstWouldAddEndPuncttrue
\mciteSetBstMidEndSepPunct{\mcitedefaultmidpunct}
{\mcitedefaultendpunct}{\mcitedefaultseppunct}\relax
\EndOfBibitem
\bibitem[Miki \emph{et~al.}(1998)Miki, Sakamoto, and Sakamoto]{Miki1998312}
K.~Miki, K.~Sakamoto and T.~Sakamoto, \emph{Surf. Sci.}, 1998, \textbf{406},
  312--327\relax
\mciteBstWouldAddEndPuncttrue
\mciteSetBstMidEndSepPunct{\mcitedefaultmidpunct}
{\mcitedefaultendpunct}{\mcitedefaultseppunct}\relax
\EndOfBibitem
\bibitem[Ravel and Newville(2005)]{Ravel:ph5155}
B.~Ravel and M.~Newville, \emph{Journal of Synchrotron Radiation}, 2005,
  \textbf{12}, 537--541\relax
\mciteBstWouldAddEndPuncttrue
\mciteSetBstMidEndSepPunct{\mcitedefaultmidpunct}
{\mcitedefaultendpunct}{\mcitedefaultseppunct}\relax
\EndOfBibitem
\bibitem[Ankudinov \emph{et~al.}(1998)Ankudinov, Ravel, Rehr, and
  Conradson]{PhysRevB.58.7565}
A.~L. Ankudinov, B.~Ravel, J.~J. Rehr and S.~D. Conradson, \emph{Phys. Rev. B},
  1998, \textbf{58}, 7565\relax
\mciteBstWouldAddEndPuncttrue
\mciteSetBstMidEndSepPunct{\mcitedefaultmidpunct}
{\mcitedefaultendpunct}{\mcitedefaultseppunct}\relax
\EndOfBibitem
\bibitem[Hohenberg and Kohn(1964)]{Hohenberg1964}
P.~Hohenberg and W.~Kohn, \emph{Phys. Rev.}, 1964, \textbf{136}, B864\relax
\mciteBstWouldAddEndPuncttrue
\mciteSetBstMidEndSepPunct{\mcitedefaultmidpunct}
{\mcitedefaultendpunct}{\mcitedefaultseppunct}\relax
\EndOfBibitem
\bibitem[Kohn and Sham(1965)]{Kohn1965}
W.~Kohn and L.~J. Sham, \emph{Phys. Rev.}, 1965, \textbf{140}, A1133\relax
\mciteBstWouldAddEndPuncttrue
\mciteSetBstMidEndSepPunct{\mcitedefaultmidpunct}
{\mcitedefaultendpunct}{\mcitedefaultseppunct}\relax
\EndOfBibitem
\bibitem[Kresse and Hafner(1993)]{Kresse1993}
G.~Kresse and J.~Hafner, \emph{Phys. Rev. B}, 1993, \textbf{48}, 13115\relax
\mciteBstWouldAddEndPuncttrue
\mciteSetBstMidEndSepPunct{\mcitedefaultmidpunct}
{\mcitedefaultendpunct}{\mcitedefaultseppunct}\relax
\EndOfBibitem
\bibitem[Kresse and Furthm\"uller(1996)]{Kresse1996}
G.~Kresse and J.~Furthm\"uller, \emph{Phys. Rev. B}, 1996, \textbf{54},
  11169\relax
\mciteBstWouldAddEndPuncttrue
\mciteSetBstMidEndSepPunct{\mcitedefaultmidpunct}
{\mcitedefaultendpunct}{\mcitedefaultseppunct}\relax
\EndOfBibitem
\bibitem[Kresse and Joubert(1999)]{Kresse1999}
G.~Kresse and D.~Joubert, \emph{Phys. Rev. B}, 1999, \textbf{59}, 1758\relax
\mciteBstWouldAddEndPuncttrue
\mciteSetBstMidEndSepPunct{\mcitedefaultmidpunct}
{\mcitedefaultendpunct}{\mcitedefaultseppunct}\relax
\EndOfBibitem
\bibitem[Perdew \emph{et~al.}(1996)Perdew, Burke, and Ernzerhof]{Perdew1996}
J.~Perdew, K.~Burke and M.~Ernzerhof, \emph{Phys. Rev. Lett.}, 1996,
  \textbf{77}, 3865\relax
\mciteBstWouldAddEndPuncttrue
\mciteSetBstMidEndSepPunct{\mcitedefaultmidpunct}
{\mcitedefaultendpunct}{\mcitedefaultseppunct}\relax
\EndOfBibitem
\bibitem[Matsuhata \emph{et~al.}(2004)Matsuhata, Sakamoto, and Miki]{Matsuhata}
H.~Matsuhata, K.~Sakamoto and K.~Miki, \emph{J. Electron Microsc (Tokyo)},
  2004, \textbf{53}, 325--337\relax
\mciteBstWouldAddEndPuncttrue
\mciteSetBstMidEndSepPunct{\mcitedefaultmidpunct}
{\mcitedefaultendpunct}{\mcitedefaultseppunct}\relax
\EndOfBibitem
\bibitem[Wolska \emph{et~al.}(2007)Wolska, Lawniczak-Jablonska, Klepka,
  Walczak, and Misiuk]{Wolska2007}
A.~Wolska, K.~Lawniczak-Jablonska, M.~Klepka, M.~S. Walczak and A.~Misiuk,
  \emph{Phys. Rev. B}, 2007, \textbf{75}, 113201\relax
\mciteBstWouldAddEndPuncttrue
\mciteSetBstMidEndSepPunct{\mcitedefaultmidpunct}
{\mcitedefaultendpunct}{\mcitedefaultseppunct}\relax
\EndOfBibitem
\bibitem[Ye \emph{et~al.}(2009)Ye, Jiang, Liu, Yao, Pan, Oyanagi, Sun, Yan, and
  Wei]{Ye_co}
J.~Ye, Y.~Jiang, Q.~Liu, T.~Yao, Z.~Pan, H.~Oyanagi, Z.~Sun, W.~Yan and S.~Wei,
  \emph{J. Appl. Phys.}, 2009, \textbf{106}, 103517\relax
\mciteBstWouldAddEndPuncttrue
\mciteSetBstMidEndSepPunct{\mcitedefaultmidpunct}
{\mcitedefaultendpunct}{\mcitedefaultseppunct}\relax
\EndOfBibitem
\bibitem[Nagaosa \emph{et~al.}(2010)Nagaosa, Sinova, Onoda, MacDonald, and
  Ong]{RevModPhys.82.1539}
N.~Nagaosa, J.~Sinova, S.~Onoda, A.~H. MacDonald and N.~P. Ong, \emph{Rev. Mod.
  Phys.}, 2010, \textbf{82}, 1539\relax
\mciteBstWouldAddEndPuncttrue
\mciteSetBstMidEndSepPunct{\mcitedefaultmidpunct}
{\mcitedefaultendpunct}{\mcitedefaultseppunct}\relax
\EndOfBibitem
\bibitem[Dietl \emph{et~al.}(2001)Dietl, Ohno, and
  Matsukura]{PhysRevB.63.195205}
T.~Dietl, H.~Ohno and F.~Matsukura, \emph{Phys. Rev. B}, 2001, \textbf{63},
  195205\relax
\mciteBstWouldAddEndPuncttrue
\mciteSetBstMidEndSepPunct{\mcitedefaultmidpunct}
{\mcitedefaultendpunct}{\mcitedefaultseppunct}\relax
\EndOfBibitem
\end{mcitethebibliography}
\bibliographystyle{rsc} 

\end{document}